\documentclass{ifacconf}

\usepackage{natbib}        

\usepackage[usenames,dvipsnames,svgnames,table]{xcolor}

\usepackage{amssymb}
\usepackage{amsmath,mathtools}
\usepackage{framed}
\usepackage{bm} 

\newcommand{\mrm}{\mathrm}

\newcommand{\pr}[1]{\left(#1\right)}
\newcommand{\sr}[1]{\left[#1\right]}

\newcommand{\add}[1]{#1}

\usepackage{physics}

\DeclareMathOperator*{\argmin}{arg\,min}

\usepackage{graphicx}

\usepackage{textcomp}
\usepackage{siunitx}
\usepackage{gensymb}
\usepackage{nicefrac}

\usepackage[normalem]{ulem}
\usepackage[makeroom]{cancel}
\newcommand{\stkout}[1]{\ifmmode\text{\sout{\ensuremath{#1}}}\else\sout{#1}\fi}

\usepackage{comment}
\usepackage{multirow}
\usepackage{makecell}
\usepackage{longtable}

\newcounter{tableeqn}[table]

\newcounter{tablesubeqn}[tableeqn]

\newcommand{\Nr}[0]{N_{r}}
\newcommand{\Vth}[0]{V_{\mathrm{th}}}
\newcommand{\Rsi}[0]{R_{\mathrm{s},i}}
\newcommand{\Rsp}[0]{R_{\mathrm{s,p}}}
\newcommand{\Rsn}[0]{R_{\mathrm{s,n}}}
\newcommand{\Ajs}[1]{A_{#1}^{\mathrm{sphere}}}

\newcommand{\FVjs}[1]{\mrm{FV}_{#1}}
\newcommand{\CVjs}[1]{\mrm{CV}_{#1}}
\newcommand{\vcsi}[0]{\vb{c}_{\mathrm{s},i}}

\newcommand{\ce}[0]{c_{\mathrm{e}}}
\newcommand{\csp}[0]{c_{\mathrm{s,p}}}
\newcommand{\csn}[0]{c_{\mathrm{s,n}}}
\newcommand{\csi}[0]{c_{\mathrm{s},i}}

\newcommand{\vbcsi}[0]{\bar{\vb{c}}_{\mathrm{s},i}}

\newcommand{\tsn}[0]{\theta_{\mathrm{s},\mathrm{n}}}
\newcommand{\tsp}[0]{\theta_{\mathrm{s},\mathrm{p}}}

\newcommand{\Dsi}[0]{D_{\mathrm{s},i}}

\newcommand{\csij}[1]{c_{\mathrm{s},i,#1}}
\newcommand{\Ni}[0]{N_{i}}
\newcommand{\Nij}[1]{N_{i,#1}}
\newcommand{\vNi}[0]{\vb{N}_{i}}

\newcommand{\bcsij}[1]{\bar{c}_{\mathrm{s},i,#1}}

\newcommand{\Dr}[0]{\Delta r}

\newcommand{\Iapp}{I_{\mathrm{app}}}

\begin{document}
\begin{frontmatter}

\title{Comparing Mass-Preserving Numerical Methods for the Lithium-Ion Battery Single Particle Model} 


\author[First]{Joseph N. E. Lucero} 
\author[Second]{Le Xu} 
\author[Second]{Simona Onori}

\address[First]{Dept. of Chemistry -- Stanford University, 
   Stanford, CA 94305 USA}
\address[Second]{Dept. of Energy Sciences and Engineering -- Stanford University, 
   Stanford, CA 94305 USA.}

\begin{abstract}                
    The single particle model (SPM) is a \add{reduced} electrochemical model that holds promise for applications in battery management systems 
    due to its ability to accurately capture battery dynamics; 
    however, the numerical discretization of the SPM requires careful consideration to ensure numerical stability and accuracy. 
    In this paper, we present a comparative study of two mass-preserving numerical schemes for the SPM: 
    the finite volume method and the control volume method. 
    Using numerical \add{simulations}, we \add{systematically} evaluate the performance of these schemes\add{, after 
    independently calibrating the SPM discretized with each scheme to experimental data,} 
    and find a tradeoff between accuracy \add{(quantified by voltage root-mean-square error)} and computational time. 
    Our findings provide insights into the selection of numerical schemes for the SPM, contributing to the advancement of battery modeling and simulation techniques.
\end{abstract}


\end{frontmatter}

\section{Introduction}

Lithium-ion batteries (LIBs) have revolutionized energy storage technology, enabling the widespread adoption of portable electronics, electric vehicles, 
and grid-scale energy storage systems. 
As demand for higher energy density, faster charging rates, and longer cycle life continue to grow, accurate modeling of LIB behavior 
become\add{s} imperative for optimizing battery design and management~\citep{rahn2013battery}. 
Specifically, in order \add{to increase} modern battery management systems\add{'} (BMS) ability to ensure maximal battery performance, 
efficiency, and ensure safe operation of battery systems, the BMS \add{may benefit from more} accurate models of the internal 
battery dynamics during operation.
Among various electrochemical models used for lithium-ion batteries, the single particle model (SPM) offers a balance between 
computational efficiency and accuracy, making \add{this model} a popular choice for simulating LIB dynamics in 
low ($\lesssim 1C$) C-rate applications~\citep{Guo2011}.
The SPM, as well as its augmented version the enhanced SPM (SPM with electrolyte \add{dynamics}), 
are also being investigated for their effectiveness for \add{on-board} state estimation~\citep{Allam2020}.

To obtain a solution of the SPM, which describes the spatiotemporal evolution of the lithium concentration within each of the battery electrodes 
during operation, the governing partial differential equations (PDEs) must be discretized. 
The choice of numerical scheme for discretization of a PDE can strongly influence the accuracy, stability, and computational cost 
of the simulations~\citep{iserles2009first}, as well as the observability of the discretized model that is important 
in estimation applications~\citep{Allam2021b}. 
Moreover, since the governing equation of lithium mass transport corresponds to a continuity equation, the total lithium mass must be 
conserved for all time\footnote{In lithium-ion batteries there are aging mechanisms 
that will cause the total (cyclable) lithium mass to decrease \add{over time.}
We neglect the modeling of such mechanisms in this work 
but \add{we} emphasize that if the numerical scheme chosen does not inherently conserve mass then it becomes difficult to attribute 
a changing total lithium mass to aging mechanisms vs. numerical error.}.
As such, numerical schemes that are used to numerical\add{ly} integrate the mass transport equation must also ensure that the 
total amount of lithium within the battery remains constant over time.

\add{A commonly used method} to discretize PDEs is known as the finite difference method (FDM); 
however, a drawback of this numerical scheme is that its structure results in a lack of mass conservation when applied to mass transport equations. 
This was demonstrated by~\cite{FordVersypt2014} when they applied FDM to a spherical diffusion equation. 
They examined different variations of FDM discretization and found that some discretizations results in numerically unstable schemes that do not preserve mass.
In the context of lithium-ion batteries, the FDM has historically \add{been} used by many to discretize the spherical diffusion equations of the SPM; 
however, the paper by~\cite{Urisanga2015} verifies the need to modify the standard FDM to address the lack of mass conservation in this case 
and proposes an alternat\add{ive} collocation-based solution.
More recently, systematic work by~\cite{Xu2023a} shows that discretizing the SPM using FDM and applying a periodic current input results in a violation of mass conservation over multiple cycles. 
In contrast, a different discretization scheme known as the ``finite volume method'' (FVM) is demonstrated to conserve mass by design 
\add{and confirmed in simulation}.

A complication of the FVM is that it does not give a direct value of the lithium surface concentration (see Sec.~\ref{sec:fvm_method_deriv}), which is needed to compute quantities such as the cell voltage. 
To obtain this, the FVM method is additionally furnished with an extrapolation method. 
Previous codes~\citep{Sulzer2021a} have used linear or quadratic Lagrange extrapolation methods that use the concentration values of the finite volumes closest to the particle surface to determine the surface concentration. 
In~\cite{Xu2023a} however, it is shown that a Hermite polynomial-based extrapolation method performs much better than the Lagrange extrapolation method at providing a surface concentration that is consistent with experimental voltage across varying input currents.

Alternatively, a different method to address the drawback of a direct (i.e. without extrapolation) estimate of the surface concentration is to 
discretize the SPM using the ``control volume method'' (CVM), initially proposed by~\cite{Zeng2013a}. 
This numerical discretization method seeks to provide the surface concentration directly without extrapolation while ensuring that 
the mass conservation property of the numerical scheme is preserved.

In this paper, we focus on comparing these two mass-preserving numerical schemes: the FVM and the CVM. 
\add{While comparisons of these two schemes has been performed previously, these studies focused on  
cases of models with non-calibrated parameters and for constant (dis)charge.}
\add{Our primary contribution is the systematic comparison of these discretization schemes when applied to the SPM model,}
\add{discretized with each scheme, and independently calibrated to experimental data.
We find that each scheme requires substantially different parameters in order to recapitulate the experimental observations.}
\add{We thus aim to re-}assess the performance of the FVM and CVM schemes in accurately capturing the transient behavior of 
lithium-ion batteries (voltage and state-of-charge) under dynamic input current profiles and the associated tradeoffs.
By comparing the numerical results obtained from these schemes against experimental data, 
we seek to provide insights into their suitability for different battery applications and simulation scenarios. 

\section{Single Particle Model}

The SPM provides a simplified yet effective framework for understanding the electrochemical processes occurring within the LIB
during operation. 
Specifically, the SPM assumes that each electrode within the battery is comprised of particles that all have the same shape and properties. 
For simplicity, the traditional SPM takes the particles to be spherical.  
Given this assumption, we can ascertain the behavior of lithium within all of the electrode particles by tracking the lithium concentration dynamics 
in just one of the particles. 
Therefore, the negative and positive electrodes $i=\{\mrm{n},\mrm{p}\}$ can be represented by a single spherical particle with radius $\Rsi$.

The lithium concentration
$\csi(r, t)$ within the bulk of a given electrode particle is assumed to 
\add{be radially symmetric and}
evolve\add{s} via Fick's law of diffusion~\eqref{eq:ficks_law}\add{,} characterized by a diffusion constant $\Dsi$.
At the surface of the electrode particle, the intercalation/de-intercalation reactions are assumed to be governed by the Butler-Volmer equation~\eqref{eq:overpotential}, from which we obtain the electrode overpotential $\eta_i$, 
given the surface concentration of the electrode particle $\csi^{\mrm{surf}}(t) = \csi(r=\Rsi,t)$, the applied current $\Iapp$\add{,} 
and the thermal voltage $V_{\mrm{th}} = RT/F$ where $R$ is the universal gas constant, $T$ is the battery temperature and $F$ is Faraday's constant. 
The state-of-charge (SoC) of each of the electrodes is determined by the bulk concentration of lithium in that electrode. 
All other lithium concentration dynamics, particularly those that would occur in the electrolyte, are considered neglible: 
The lithium concentration in the electrolyte is assumed to be fixed at its average value $\ce^{\mrm{avg}} = 1000$.
The SPM model governing equations are summarized in Table~\ref{tbl:SPM}.
\begin{table}[!htbp]
	\caption{Governing equations of the SPM}
	\label{tbl:SPM}
	\centering
    \large
	\setlength{\tabcolsep}{1mm}{
		\resizebox{1.0\linewidth}{!}{
			\begin{tabular}{ll}
				\Xhline{1.5 pt}
				Variable & Equation  \\ \hline
				Fick's Law of Diffusion & \parbox{10cm}{ \begin{equation} \label{eq:ficks_law} \begin{array}{*{20}{c}} {\displaystyle \Ni(r,t)=-\Dsi \pdv{\csi}{r},}&{i = \{\mrm{n},\mrm{p}\}}  \end{array} \end{equation}} \\
				Mass transport in solid phase & \parbox{10cm}{ \begin{equation} \label{eq:mass_transport} \begin{array}{*{20}{c}} {\displaystyle \pdv{\csi}{t} = -\div{\Ni}}  \end{array} \end{equation}} \\
                Boundary conditions &  \parbox{10cm}{ \begin{equation} \label{eq:boundary_cond}
						\begin{array}{*{20}{c}}
							{\Ni(r=0,t) = 0;}&{\Ni(r=\Rsi,t) = \dfrac{{{I_{\mrm{app}}}g_{i}}}{{{a_{\mrm{s},i}}{A_{\mrm{cell}}}{L_i}F}}} \\ 
							{{a_{\mrm{s},i}} = \dfrac{3}{\Rsi}{\varepsilon_i},}&{g_{i} = \left\{ {\begin{array}{*{20}{c}}
										{\begin{array}{*{20}{c}}
												{-1,}&{i = \mrm{p}}
										\end{array}} \\ 
										{\begin{array}{*{20}{c}}
												{1,}&{i = \mrm{n}}
										\end{array}} 
								\end{array}} \right.} 
				\end{array} \end{equation}}  \\
				Electrode overpotential &  \parbox{10cm}{ \begin{equation} \label{eq:overpotential} 
                            \begin{array}{*{20}{c}} {\eta_{i} = 2\cdot\Vth\cdot\sinh^{-1}\pr{{\dfrac{{I_{\mrm{app}}}g_{i}}{{2{a_{s,i}}A{L_i}Fj_{0,i}}}}},} \end{array} 
                        \end{equation}} \\
				Exchange flux density & \parbox{10cm}{ \begin{equation} \label{eq:exchange_flux}
						j_{0,i} = k_{0,i}\sqrt{\ce^{\mrm{avg}}\pr{\frac{\csi^{\mrm{surf}}}{\csi^{\mrm{max}}}}\pr{1 - \frac{\csi^{\mrm{surf}}}{\csi^{\mrm{max}}}}} \end{equation}} \\
				Cell voltage & \parbox{10cm}{ \begin{equation} \label{eq:voltage_eqn}
						V_{\mrm{cell}} = U_{\mrm{p}}\pr{\frac{\csp^{\mrm{surf}}}{\csp^{\mrm{max}}}} - U_{\mrm{n}}\pr{\frac{\csn^{\mrm{surf}}}{\csn^{\mrm{max}}}} + \eta_{\mrm{p}} - \eta_{\mrm{n}} - R_{\ell} \cdot I_{\mrm{app}} \end{equation}} \\
				State-of-Charge & \parbox{10cm}{ \begin{equation} \label{eq:soc_eqn}
						\begin{gathered}
							\tsp^{\mrm{bulk}} = \frac{1}{\frac{4}{3}\pi \Rsp^3}\int_{V_{\mrm{n}}^{\mrm{sphere}}} \pr{\frac{\csp}{\csp^{\mrm{max}}}}\dd{V} \hfill \\
							\tsn^{\mrm{bulk}} = \frac{1}{\frac{4}{3}\pi \Rsn^3}\int_{V_{\mrm{p}}^{\mrm{sphere}}} \pr{\frac{\csn}{\csn^{\mrm{max}}}}\dd{V} \hfill \\ 
							\begin{array}{*{20}{c}}
								{\mrm{SoC}_{\mrm{p}} = \dfrac{{\tsp^{0\%} - \tsp^{\mrm{bulk}}}}{{\tsp^{0\%} - \tsp^{100\%}}},}&{\mrm{SoC}_{\mrm{n}} = \dfrac{{\tsn^{\mrm{bulk}} - \tsn^{0\%}}}{{\tsn^{100\%} - \tsn^{0\%}}}} 
							\end{array} \hfill \\
				\end{gathered} \end{equation}} \\ \Xhline{1.5 pt}
	\end{tabular}}}
\end{table}

\subsection{Numerical methods}        

To solve the dynamics of the SPM model, we must discretize the mass transport equation~\eqref{eq:mass_transport}. 
For each electrode particle $i$, we define a set of $\Nr$ nodes for which we discretize the radial coordinate $r$,
{\small
\begin{align}
    \label{eq:grid_defn}
    r_{i,j} = (j-1)\Dr_{i}\ ,\ j\ \in\ [1, 2, \dots, \Nr]\ .
\end{align}
}%
These nodes are spaced by a distance,
{\small
 \begin{align}
    \Dr_{i} = \frac{\Rsi}{\Nr-1}\ .
\end{align}
}%
In this work, although the radius $\Rsi$ may be different for each electrode, 
we assume that the number of nodes $\Nr$ is the same for both electrodes.
Half-distances between nodes will also be commonly referred to throughout this work. 
For notational convenience, we define
{\small
\begin{subequations}
    \begin{align}
        r_{i,j} + \frac{\Dr_{i}}{2} &= (j-1)\Dr_{i} + \frac{\Dr_{i}}{2} \\
        &= \pr{j - \frac{1}{2}}\Dr_{i} \\
        &\equiv r_{i,j+1/2}\ .
    \end{align}
\end{subequations}
}

\subsubsection{Finite Volume Method} 
\label{sec:fvm_method_deriv}

To discretize the mass transport equation~\eqref{eq:mass_transport} using the finite volume method
we define the $j$th ``finite volume'' for each electrode $i$ to be the spherical shell with $r\in[r_{i,j},r_{i,j+1}]$ centered on the point $r_{i,j+1/2}$, 
having a volume of
{\small
\begin{align}
    \FVjs{j+1/2} = \frac{4\pi}{3}\pr{r_{i,j+1}^{3}-r_{i,j}^{3}}\ .
\end{align}
}%
As the edges of the finite volumes are the nodes defined in~\eqref{eq:grid_defn}, there are a total of $\Nr-1$ finite volumes.  
For each of these small volumes, integrating~\eqref{eq:mass_transport}, applying the divergence theorem, and using~\eqref{eq:ficks_law}, we obtain
{\small
\begin{align}
    \int_{\FVjs{j+1/2}}\pdv{\csi}{t}&\dd{V} \label{eq:FVM_weak_form}\\
    &\hspace{-2.5cm}= \int_{\Ajs{j+1}}\Dsi\pr{\pdv{\csi}{r}}\dd{A} - \int_{\Ajs{j}}\Dsi\pr{\pdv{\csi}{r}}\dd{A}\ ,\nonumber
\end{align}
}%
where the surface area $\Ajs{j} = 4\pi r_{j}^{2}$. 
Equation~\eqref{eq:FVM_weak_form} can be discretized into a system of $\Nr-1$ ordinary differential equations (ODEs) with a state vector $\vbcsi^{\mrm{FVM}}$
whose elements represents the average concentration within the $j$th finite volume,
{\small
\begin{align}
    \vbcsi^{\mrm{FVM}} = \begin{bmatrix} \bcsij{3/2} & \bcsij{5/2} & \cdots & \bcsij{\Nr-3/2} & \bcsij{\Nr-1/2} \end{bmatrix}^{\mrm T}\ ,
\end{align}
}%
and obeys the following state-space equation,
\begin{align}
    \dv{\vcsi^{\mrm{FVM}}}{t} = A^{\mrm{FVM}}\vcsi^{\mrm{FVM}} + B^{\mrm{FVM}}\Iapp\ .\label{eq:fvm_ss_eqn}
\end{align}
Here, $\pr{A^{\mrm{FVM}}}_{\Nr-1\times\Nr-1}$ is a tridiagonal matrix and $\pr{B^{\mrm{FVM}}}_{\Nr-1\times 1}$ is a vector.
For the specific forms of these matrices and vectors and their derivation, we direct the reader to reference~\cite[Eqs. 38-41]{Xu2023a}.
We emphasize that, by its structure, this method is mass conserving.

To compute the cell voltage~\eqref{eq:voltage_eqn} the surface concentration $\csi^{\mrm{surf}}$ is needed. 
As the finite volume method does not yield the surface concentration directly, it is common to use linear extrapolation to obtain 
the surface concentration given the concentrations of the two finite volumes closest to the particle surface.
However, recently~\cite{Xu2023a} also showed that using a Hermite polynomial-based extrapolation scheme for the surface 
concentration results in greater accuracy in predicting the output voltage of the battery cell.
As such, we will be using this method in this work.

\subsubsection{Control Volume Method} 
\label{sec:cvm_method_deriv}

To discretize the mass transport equation~\eqref{eq:mass_transport} using the control volume method,
we define the $j$th ``control volume'' to be the spherical shell with $r\in[r_{i,j-1/2},r_{i,j+1/2}]$
centered on the (non-boundary) grid point $r_{i,j}$, with volume
{\small
\begin{align}
    \CVjs{j} = \frac{4\pi}{3}\pr{r_{j+1/2}^{3}-r_{j-1/2}^{3}}\ . 
\end{align}
}%
Similar to the FVM method derivation, we integrate both sides of~\eqref{eq:mass_transport} over each of these small volumes. 
Applying the divergence theorem on the right-hand side we obtain,
{\small
\begin{align}
    \label{eq:cvm_discrete}
    \int_{\CVjs{j}}\pdv{\csi}{t}&\dd{V} = -\int_{\Ajs{j+1/2}}\Ni(r,t)\dd{A} + \int_{\Ajs{j-1/2}}\Nr(r,t)\dd{A}\ .
\end{align}
}%
where the surface area $\Ajs{j+1/2} = 4\pi r_{j+1/2}^{2}$. 
Similar to the finite volume method,~\eqref{eq:cvm_discrete} can be discretized into a system of $\Nr$ ODEs for the average concentration 
within the control volume $\vbcsi^{\mrm{CVM}}$ where each element represents the average concentration of the $j$th control volume,
{\small
\begin{align}
    \vbcsi^{\mrm{CVM}} = \begin{bmatrix} \bcsij{1} & \bcsij{2} & \cdots & \bcsij{\Nr-1} & \csij{\Nr} \end{bmatrix}^{\mrm T}\ ,
\end{align}
}%
and is governed by the state-space representation,
{\small
\begin{align}
    M^{\mrm{CVM}}\dv{\vbcsi^{\mrm{CVM}}}{t} = A^{\mrm{CVM}}\vNi\ ,\label{eq:cvm_ss_eqn}
\end{align}
}%
where $\pr{M^{\mrm{CVM}}}_{\Nr\times\Nr}$ and $\pr{A^{\mrm{CVM}}}_{\Nr\times\Nr}$ are both tridiagonal matrices and the vector of fluxes is
{\small
\begin{align}
    \vNi = \begin{bmatrix} \Nij{3/2} & \Nij{5/2} & \cdots & \Nij{\Nr-1/2} & \frac{\Iapp g_{i}}{a_{\mrm{s},i}A_{\mrm{cell}}L_{i}F} \end{bmatrix}_{1\times{\Nr}}^{\mrm T}\ ,
\end{align}
}%
with 
{\small
\begin{align}
    \Nij{j+1/2} = \Dsi\frac{\bcsij{j+1}-\bcsij{j}}{\Dr}\ .
\end{align}
}%
\add{Importantly, in this scheme, the surface concentration is part of the state vector.}
For explicit forms of the matrices $M^{\mrm{CVM}}$ and $A^{\mrm{CVM}}$ we direct the reader to~\cite[Eqs. 18-20]{Zeng2013a}.
By construction, as detailed in the original publication, this method is mass conserving. 

\subsection{Choice of integrator}

To integrate the state-space equations~\eqref{eq:fvm_ss_eqn} and~\eqref{eq:cvm_ss_eqn}, we use the built-in MATLAB ODE integrator \verb|ode15s|. 
This is a variable-step variable-order integration method in MATLAB that can handle stiff systems of ODEs and state-space equations 
involving mass matrices such as in~\eqref{eq:cvm_ss_eqn}. 
We set the integrator absolute tolerance \verb|abstol| to be $10^{-11}$ and the relative tolerance \verb|reltol| to be $10^{-8}$.

\section{Results}

To test the effectiveness of both of these discretization methods when applied to the SPM, 
we compare their ability to describe the behavior of an LG INR21700-M50T NMC/Gr cell where the data was initially collected by~\cite{Pozzato2022}. 
This particular cell was also experimentally characterized earlier by~\cite{Chen2020}.
From these two studies, the values of model parameters that are intrinsic to the M50T cell are borrowed and listed in Table~\ref{tbl:borrowed_parameters}.
\begin{table}[!htbp]
    \centering
    \renewcommand{\arraystretch}{1.2}
    \caption{Borrowed parameters for both discretization methods.}
    \label{tbl:borrowed_parameters}
    \begin{tabular}{|c|c|c|}
    \hline
        Borrowed Parameter                          & Value  \\ \hline
        $\theta_{\mrm{s,n}}^{100\%}\ (-)$           & 0.9343 \\ \hline
        $\theta_{\mrm{s,p}}^{100\%}\ (-)$           & 0.2711 \\ \hline
        $\theta_{\mrm{s,n}}^{0\%}\ (-)$             & 0.0204 \\ \hline
        $\theta_{\mrm{s,p}}^{0\%}\ (-)$             & 0.8536 \\ \hline
        $\csn^{\mrm{max}}\ (\mrm{mol}/\mrm{m}^{3})$ & 29583  \\ \hline
        $\csp^{\mrm{max}}\ (\mrm{mol}/\mrm{m}^{3})$ & 51765  \\ \hline
        $\ce^{\mrm{avg}}\ (-)$                      & 1000   \\ \hline
        $L_{\mrm{n}}\ (\mu\mrm{m})$                 & 85.2   \\ \hline
        $L_{\mrm{p}}\ (\mu\mrm{m})$                 & 75.6   \\ \hline
        $A_{\mrm{cell}}\ (\mrm{cm}^{2})$            & 1126.7 \\ \hline
        $R_{\ell}\ (\mrm{m}\Omega)$                 & 29     \\ \hline
    \end{tabular}
\end{table}
The rest of the SPM model parameters are calibrated to the collected experimental data as explained in the next section.

\subsection{Model Calibration -- Procedure}
We first re-calibrate both methods \add{independently} to experimental data. 
Specifically, for each method, we seek the vector of model parameters that are related to the geometry 
and dynamics assumed by the SPM model,
{\small
\begin{align}
    \bm{\lambda}_{\mrm{iden}} = 
    \begin{bmatrix} \Rsn & \Rsp & \varepsilon_{\mrm{n}} & \varepsilon_{\mrm{p}} & D_{\mrm{s,n}} & D_{\mrm{s,p}} & k_{0,\mrm{n}} & k_{0,\mrm{p}}\end{bmatrix}^{\mrm T}\ ,
\end{align}
}%
that minimizes the objective function $J$,
{\small
\begin{align}
   \bm{\lambda}_{\mrm{iden}}^{*} &= \argmin_{\bm{\lambda}_{\mrm{iden}}^{\mrm{min}}\le\bm{\lambda}_{\mrm{iden}}\le\bm{\lambda}_{\mrm{iden}}^{\mrm{max}}} 
   \overbrace{\sr{J_{V} + J_{\mrm{SoC}_{\mrm{p}}} + J_{\mrm{SoC}_{\mrm{n}}}}}^{= J}\ ,\label{eq:min_objective}
\end{align}
}%
that is the sum of the following three terms, 
{\small
\begin{subequations}
    \begin{align}
       J_{V} &= \sqrt{\frac{1}{M}\sum_{k=1}^{M}\pr{1-\frac{V_{\mrm{cell}}^{\mrm{sim}}(\bm{\lambda}_{\mrm{iden}},t_k,\Iapp)}{V_{\mrm{cell}}^{\mrm{exp}}(t_k)}}^{2}}\\
       J_{\mrm{SoC}_{\mrm{p}}} &= \sqrt{\frac{1}{M}\sum_{k=1}^{M}\pr{1-\frac{\mrm{SoC}_{\mrm{p}}^{\mrm{sim}}(\bm{\lambda}_{\mrm{iden}},t_k,\Iapp)}{\mrm{SoC}^{\mrm{exp}}(t_k)}}^{2}}\\
       J_{\mrm{SoC}_{\mrm{n}}} &= \sqrt{\frac{1}{M}\sum_{k=1}^{M}\pr{1-\frac{\mrm{SoC}_{\mrm{n}}^{\mrm{sim}}(\bm{\lambda}_{\mrm{iden}},t_k,\Iapp)}{\mrm{SoC}^{\mrm{exp}}(t_k)}}^{2}}\ .
    \end{align}
\end{subequations}
}%
The objective function $J$ measures the sum of the root-mean-square error between the experimental and model voltage, as well as positive and negative electrode SoC's, respectively.
Here, $M$ denotes the total number of time points for which we have experimental data.
The vectors $\bm{\lambda}_{\mrm{iden}}^{\mrm{min}}$, $\bm{\lambda}_{\mrm{iden}}^{\mrm{max}}$ denote the lower and upper bounds for the parameter vector
with elements \add{(in SI units)}
{\small
\hspace*{-1cm}
\begin{subequations}
    \begin{align}
        \bm{\lambda}_{\mrm{iden}}^{\mrm{min}} &= 
        \begin{bmatrix}  10^{-6} & 10^{-6} & 0.6 & 0.6 & 10^{-17} & 10^{-17} & 10^{-7} & 10^{-7}\end{bmatrix}^{\mrm T}\\
        \bm{\lambda}_{\mrm{iden}}^{\mrm{max}} &= 
        \begin{bmatrix} 1.2\cdot10^{-5} & 1.2\cdot10^{-5} & 0.8 & 0.8 & 10^{-10} & 10^{-10} & 10^{-2} & 10^{-2}\end{bmatrix}^{\mrm T}\ .
    \end{align}
\end{subequations}
}%
The parameter value bounds are chosen to ensure that the parameters are within reasonable physical ranges that are roughly consistent with previous measurements outlined in~\cite{Chen2020}.
For this process, we collect experimental output voltage data $V_{\mrm{cell}}^{\mrm{exp}}$ from a cell placed galvanostatically under a hybrid pulse power characterization (HPPC) input as discussed in~\cite{Pozzato2022}.
The experimental state-of-charge $\mrm{SoC}^{\mrm{exp}}$ is obtained by Coulomb Counting method given the current input $\Iapp$.

To perform the minimization~\eqref{eq:min_objective}, we use the Particle Swarm Optimization (PSO) algorithm which is part of MATLAB's Global Optimization Toolbox. 
Specifically, we set the algorithm with a \verb|SwarmSize| = 80, a \verb|SelfAdjustmentWeight| = 2, a \verb|SocialAdjustmentWeight| = 1, and \verb|MinNeighborsFraction| = 1.
Each method discretizes the radial dimension of each electrode particle with $\Nr=101$ node points \add{to ensure maximal resolution}.

\subsection{Model Calibration -- Results}



We perform the calibration and the resulting calibration errors are quantitatively summarized in Table~\ref{tbl:summary_vals}.
\begin{table}[!htbp]
    \centering
    \caption{Objective function values obtained for different input profiles for each method}
    \renewcommand{\arraystretch}{1.2}
    \label{tbl:summary_vals}
    \begin{tabular}{|c|c|c|c|c|c|}
        \hline
        Method               & Input & $J_{V}\ (\%)$ & $J_{\mrm{SoC}_{\mrm{p}}}\ (\%)$ & $J_{\mrm{SoC}_{\mrm{n}}}\ (\%)$ & $J (\%)$ \\ \hline
        \multirow{2}{*}{FVM} & HPPC  & 0.55          & 0.02                            & 0.02                            & 0.60     \\ \cline{2-6} 
                             & UDDS  & 0.45          & 0.01                            & 0.01                            & 0.48     \\ \hline
        \multirow{2}{*}{CVM} & HPPC  & 0.55          & 0.02                            & 0.02                            & 0.60     \\ \cline{2-6} 
                             & UDDS  & 0.46          & 0.01                            & 0.01                            & 0.49     \\ \hline
    \end{tabular}
\end{table}
In general, we observe a good fit of the model to experimental HPPC data using with both numerical methods, 
with an objective function value of $J < 1\%$ for both methods which corresponds to a voltage root-mean-square error (RMSE), 
{\small
\begin{align}
    \mrm{RMSE} = \sqrt{\frac{1}{M}\sum_{k=1}^{M}\pr{V_{\mrm{cell}}^{\mrm{sim}}(\bm{\lambda}_{\mrm{iden}},t_k,\Iapp)-V_{\mrm{cell}}^{\mrm{exp}}(t_k)}^{2}}\ ,
\end{align}
}%
of $19.71\ \mrm{mV}$ for the FVM method and $19.70\ \mrm{mV}$ for the CVM method.
Thus, we conclude that the performance of the FVM and the CVM on calibration is nearly equivalent. 

The model parameters obtained from the calibration process for each method are summarized in Table~\ref{tbl:parameters}.
Interestingly, the discretization method has a significant effect on the inferred parameter values.
\add{As such, it is important to independently calibrate both methods to ensure maximal accuracy and a fair comparison in their ability
to reproduce the experimental observations.} 
In particular, the diffusion coefficient $D_{\mrm{s,n}}$ and the kinetic constant for the negative electrode $k_{0,\mrm{n}}$ are 
\add{seen to be} between 1 to 3 orders of magnitude different between the two methods.
\begin{table}[!htbp]
    \centering
    \renewcommand{\arraystretch}{1.2}
    \caption{Obtained parameters for both discretization methods.}
    \label{tbl:parameters}
    \begin{tabular}{|c|c|c|}
    \hline
        Parameter                                                & FVM value            & CVM value           \\ \hline
        $\Rsn\ (\mu\mrm{m})$                                     & 10.70                & 2.87                \\ \hline
        $\Rsp\ (\mu\mrm{m})$                                     & 12                   & 6.31                \\ \hline
        $\varepsilon_{\mrm{n}}\ (-)$                             & 0.76                 & 0.76                \\ \hline
        $\varepsilon_{\mrm{p}}\ (-)$                             & 0.77                 & 0.77                \\ \hline
        $D_{\mrm{s,n}}\ (\mu\mrm{m}^{2}/\mrm{s})$                & $0.17$               & $8.19\times10^{-3}$ \\ \hline
        $D_{\mrm{s,p}}\ (\mu\mrm{m}^{2}/\mrm{s})$                & $8.19\times10^{-3}$ & $2.13\times10^{-3}$  \\ \hline
        $k_{0,\mrm{n}}\ (\mrm{mmol}/(\mrm{m}^{2}\cdot \mrm{s}))$ & $6.95\times10^{-6}$  & $3.06\times10^{-3}$ \\ \hline
        $k_{0,\mrm{p}}\ (\mrm{mmol}/(\mrm{m}^{2}\cdot \mrm{s}))$ & $2.11$  & $3.39$  \\ \hline
    \end{tabular}
\end{table}

\subsection{Model Validation}
\begin{figure}[!htbp]
    \centering
    \includegraphics[clip, width=\linewidth]{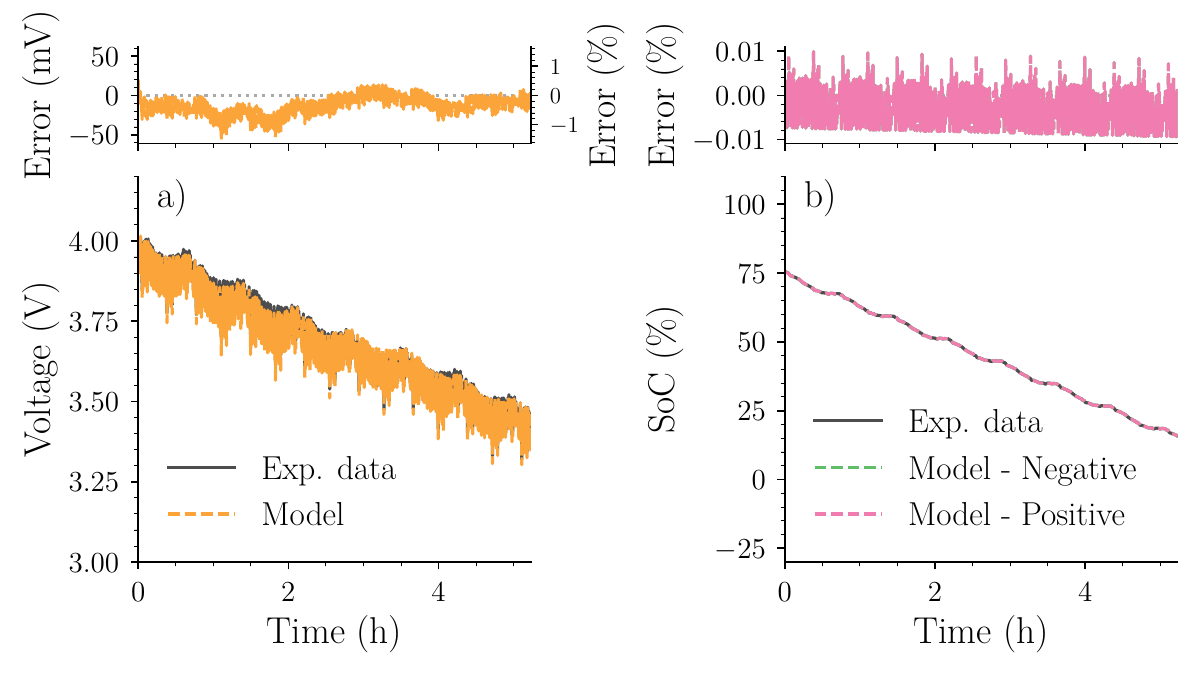}
    \caption{Validation of FVM method with Hermite polynomial extrapolation using UDDS data. 
    (a) Voltage comparison. (b) SoC comparison. Solid lines denote experimental data. Dashed lines denote model outputs.
    Top plots show the error (difference) between model output and experimental data..}
    \label{fig:FVM_UDDS_results}
\end{figure}

\begin{figure}[!htbp]
    \centering
    \includegraphics[clip, width=\linewidth]{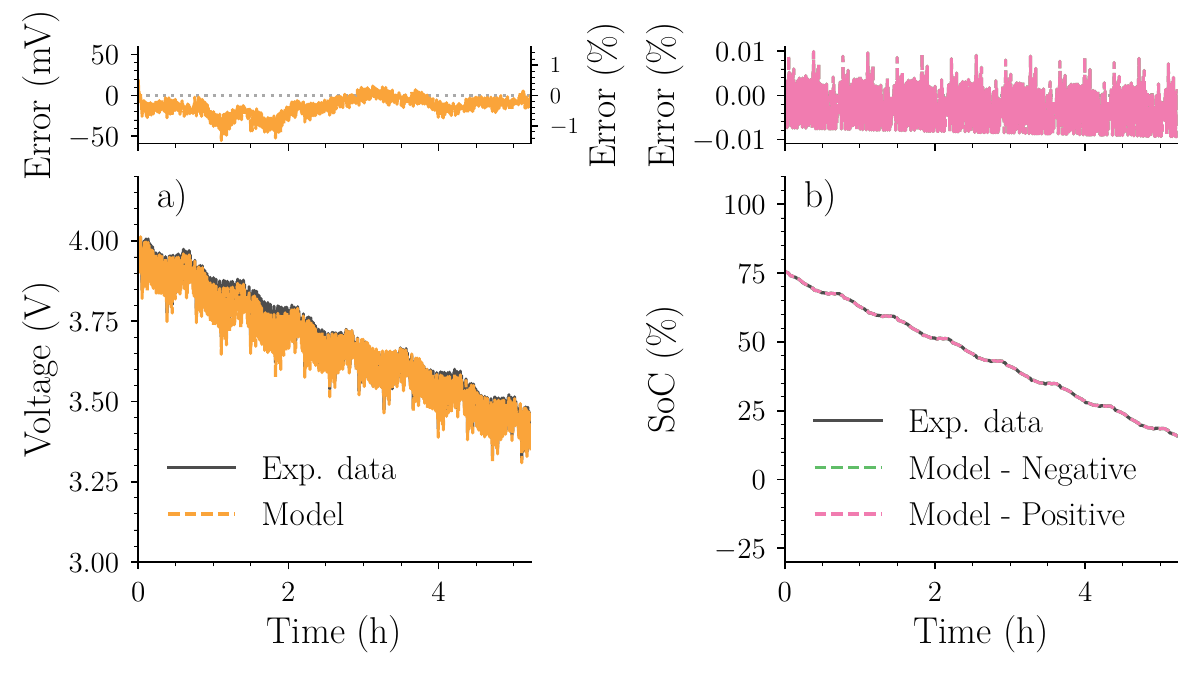}
    \caption{Validation of CVM method using UDDS data. Similar visualization as Fig.~\ref{fig:FVM_UDDS_results}.}
    \label{fig:CVM_UDDS_results}
\end{figure}

After calibrating both methods to HPPC experimental data, we now seek to validate our models using a dynamic drive cycle.
We collect experimental voltage data from a cell where an Urban Dynamometer Driving Schedule (UDDS) input current profile
has been applied.
Figures~\ref{fig:FVM_UDDS_results} and~\ref{fig:CVM_UDDS_results} compares the model outputs to the experimental data for the FVM and CVM schemes under this input, respectively. 
We find that the model discretized by FVM yields a validation RMSE of $16.83\ \mrm{mV}$ while the model discretized using CVM yields a validation RMSE of $17.25\ \mrm{mV}$.
We therefore again see that the performance of the two methods are nearly equivalent.

\subsection{Volume number variation}

As the SPM is \add{a simple} electrochemical physics-based model of the lithium-ion battery, it is particularly suited for BMS applications; 
however, in order for this model to be able to run in real-time the number of node points for each electrode must be kept \add{small}. 
Thus, having validated the performance of both models on a dynamic drive cycle profile, we now investigate how the performance of the 
two methods varies as we decrease the number of node points $\Nr$.
We emphasize that we do not perform any further calibrations here: The parameters for each method are fixed according to 
Tables~\ref{tbl:borrowed_parameters} and~\ref{tbl:parameters} for all the analysis that follows.

\begin{figure}[!htbp]
    \centering
    \includegraphics[clip, width=\linewidth]{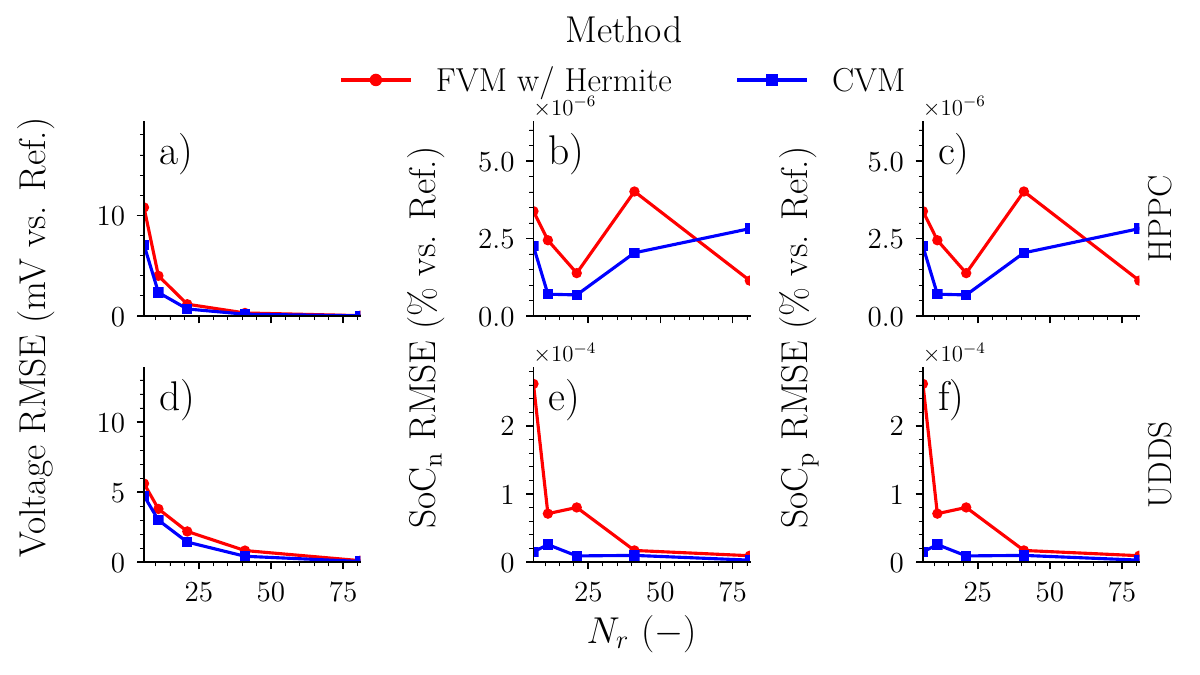}
    \caption{RMSE between model output of using a given method and reference solution. (a,d) Voltage RMSE. (b, e) Negative electrode SoC RMSE. 
    (c, f) Positive electrode SoC RMSE. (a-c) are associated with an HPPC input while (d-f) are associated with a UDDS input.}
    \label{fig:error_vs_UDDS}
\end{figure}

To measure the performance of these two methods, we use as reference the results for each discretization method with $\Nr = 101$ node points 
presented in the previous sections.
We quantify the difference between the solution generated by a \add{given} method using $\Nr = 101$ node points and the 
solution generated using a smaller number of node points.
Specifically, we scan across values of $\Nr = \{6, 11, 21, 41, 81\}$, where $\Nr = 6$ is a number of node points that is consistent 
with use in BMS-type applications. 
This comparison of the different outputs from the different methods, measured in terms of RMSE, 
is shown in Fig.~\ref{fig:error_vs_UDDS} for both HPPC and UDDS inputs.
In general, we see that the voltage RMSE error between the model output and the experimental data grows monotonically as $\Nr$ decreases. 
Interestingly, the FVM method using Hermite polynomial extrapolation diverges from the reference solution in terms of voltage faster than the CVM method. 
A similar trend holds for the state-of-charge; however, we note that the initial scale of the error for SoC is already quite small, on the order of 0.01\%.

\subsection{Perspective}

The CVM method \add{produces} an implicit equation~\eqref{eq:cvm_ss_eqn} that requires solving a linear system at every time step 
to obtain the time derivatives $\mrm{d}\vcsi/\mrm{d}t$. 
In contrast, the FVM method \add{results in} an explicit equation~\eqref{eq:fvm_ss_eqn}
that requires only a matrix-vector product and a vector addition to evaluate the time derivative. 
As such, we reason that the CVM method should be expected to be more computationally expensive relative to the FVM.
Furthermore, a fixed number $\Nr$ of node points is used to discretize the radial coordinate for both methods. 
This corresponds to integrating forward in time $\Nr-1$ coupled equations for the FVM but $\Nr$ coupled equations for the FVM. 
Thus, since CVM integrates one more equation than the SPM, it stands to reason that this would also contribute to a higher computational time for the CVM.

\begin{figure}[!htbp]
    \centering
    \includegraphics[clip, width=\linewidth]{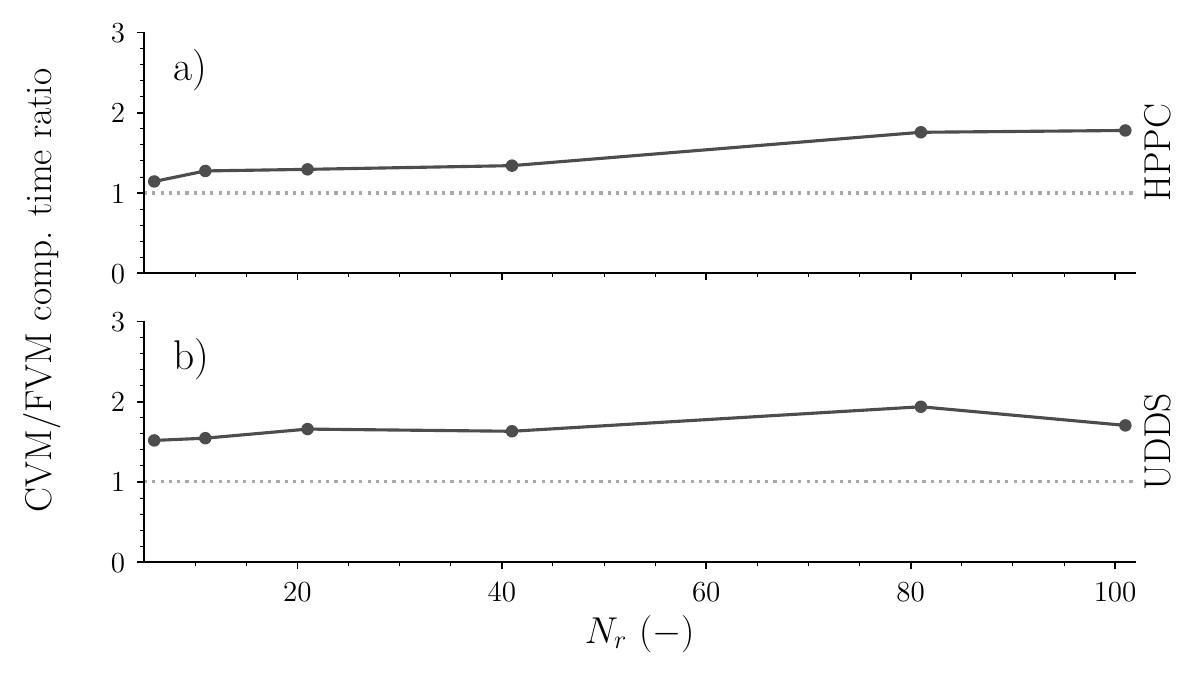}
    \caption{Ratio of CVM computational time to FVM computational time as a function of the number of node points $\Nr$ for an 
    (a) HPPC profile and (b) a UDDS profile. Error bars denote the standard error of the mean and are present but are as small as the points.}
    \label{fig:computational_time}
\end{figure}

To verify this, we pick\add{ed} a set of node point numbers $\Nr = \{6, 11, 21, 41, 81, 101\}$ and for each of these node point numbers, 
we solve\add{d} the SPM model using a particular discretization scheme, under a particular input profile, five times. 
We measure (using MATLAB's built in timer) the time \add{for} the numerical solver \add{to} return a \add{solution} and obtain 
an average run time by averaging over the five replicates. 
The ratio of CVM computation times to FVM computation time as a function of $\Nr$ is shown in 
Fig.~\ref{fig:computational_time} which confirms our expectation that the 
CVM method requires more computational effort compared to the FVM with the ratio attaining the largest value of 
1.9 for $\Nr = 81$ node points when applying the UDDS profile but generally remaining near one.

As the CVM achieves superior demonstrated performance compared to the FVM in terms of accuracy for the SPM, its less widespread adoption presents a unique opportunity for advancement in battery modeling. 
In contrast, FVM is increasingly being used in battery models and the addition of the Hermite polynomial extrapolation method increases the FVM's utility and performance over a traditional method such as FDM.
Nonetheless, as CVM is able to provide higher accuracy while maintaining mass conservation, it signifies the potential for this method to significantly elevate the fidelity of battery simulations, particularly those used in BMS; 
however, there are also challenges with this method's usage. In particular, the need for specialized knowledge in order to implement this method as well as the computational overhead are noteworthy tradeoffs. 

\section{Conclusion}

In this comparative study, we have explored the performance of two mass-preserving numerical schemes, namely the finite volume method (FVM) and the control volume method (CVM), in the context of the single particle model (SPM) for lithium-ion batteries. 
By comparing the numerical outputs of the SPM discretized using both methods, we demonstrate that both methods attain essentially equivalent performance in predicting the voltage and state-of-charge dynamics of an LG INR21700-M50T NMC/Gr cell during dynamic operation\add{; however, 
the identified model parameters for each scheme are substantially different.}
Investigating how the solution \add{accuracy} provided by each method changes as the number of node points is decreased, 
or equivalently the degree of discretization is increased, shows that CVM is more robust than FVM to the level of discretization.
This makes it an attractive choice for applications where accuracy at large discretizations is important, such as in battery management systems.

On the other hand, the increased robustness of CVM comes at increased computational cost. 
Specifically, our analysis reveals that the computational time required to numerically solve the mass transport equations of the SPM model using CVM is generally higher compared to that of FVM. 
The structure associated with the state-space equation of the CVM, requiring a linear solve at every time step, contributes to this increased computational burden.
Moreover, we also acknowledge that implementation of the FVM and CVM is not as straightforward as a direct discretization of the mass transport equations using finite-difference methods. 
The formulation of both FVM and CVM methods requires a deeper understanding of numerical techniques and may pose challenges for researchers and practitioners unfamiliar with the intricacies of PDE discretization.
Future research efforts may focus on developing efficient algorithms and computational strategies to mitigate the computational overhead associated with the CVM, thereby further enhancing its usability and applicability in real-time battery monitoring.


\begin{ack}
    The authors thank Dr. Yizhao Gao and Sung Yeon Sara Ha (Stanford Energy Science \& Engineering) for helpful discussions.
\end{ack}

\end{document}